\documentclass[12pt]{article}
\usepackage{amsmath}
\usepackage{amsfonts}
\usepackage{amssymb}
\usepackage{graphicx}
\usepackage{color}

\numberwithin{equation}{section}

\makeatletter
\newcommand{\rd}{\@ifnextchar^{\DIfF}{\DIfF^{}}}
\def\DIfF^#1{%
   \mathop{\mathrm{\mathstrut d}}%
   \nolimits^{#1}\gobblespace}
\def\gobblespace{\futurelet\diffarg\opspace}
\def\opspace{%
   \let\DiffSpace\!%
   \ifx\diffarg(%
   \let\DiffSpace\relax
   \else
   \ifx\diffarg[%
   \let\DiffSpace\relax
   \else
   \ifx\diffarg\{%
   \let\DiffSpace\relax
   \fi\fi\fi\DiffSpace}

\providecommand*{\iu}%
{\ensuremath{\mathrm{i}\,}}
\def\b{\beta}

\def\d{\delta}

\def\br{\begin{eqnarray}}
\def\er{\end{eqnarray}}
\def\be{\begin{equation}}
\def\ee{\end{equation}}
\def\nonu{\nonumber}
\def\lb{\lbrack}
\def\rb{\rbrack}
\def\({\left(}
\def\){\right)}

\relax

\def\a{\alpha}
\def\b{\beta}

\def\d{\delta}

\def\eps{\epsilon}

\def\l{\lambda}
\def\L{\Lambda}

\def\o{\over}
\def\O{\Omega}

\def\pa{\partial}

\def\s{\sigma}

\def\lie{{\cal G}}

\def\PLA#1#2#3{{\sl Phys. Lett.} {\bf #1A} (#2) #3}

\begin{document}
%\begin{titlepage}

%\vspace*{\stretch{1.0}}

   \begin{center}
 \Large\textbf{Miura and Generalized B\"acklund Transformation  for KdV Hierarchy
}
%as Reduction of Integrable $\mathbf{4}$-Boson Model}
\end{center}
\vskip 9 mm
\begin{center}
{\bf J. F. Gomes, 
A. L. Retore and A. H. Zimerman}\\
Instituto de F\'isica Te\'orica-UNESP\\
Rua Dr Bento Teobaldo Ferraz 271, Bloco II\\
01140-070, S\~ao Paulo, Brazil\\
%e-mail: jfg@ift.unesp.br\\
\end{center}

\vskip 9 mm

\abstract{Using the fact that  Miura transformation can be   expressed  in the form of gauge transformation  
connecting the KdV and mKdV equations, we  discuss the derivation  of
the B\"acklund transformation  and its Miura-gauge transformation connecting both hierarchies. }

\section{Introduction}

Integrable models are  characterized by   an infinite number of conservation laws which are responsible  for the stability of soliton solutions. 
In fact,  these  conservation laws may be regarded as hamiltonians generating time evolutions  within a multi-time space.  
Each  of these  time evolutions, in turn,   are associated to a non-linear  equation of motion and henceforth  constitute an integrable hierarchy of equations with a common set of conservation laws.
It has  become clear in the past few years that integrable hierarchies  may be systematically constructed  from a graded algebraic structure involving  affine Kac-Moody algebras.  
The equations of motion  are   constructed  systematically from  a graded  algebra     embedded within  a zero curvature representation.
The most well known example are the sine (sinh)-Gordon  and mKdV equations, both  underlined by the same  $\hat sl(2)$ affine  algebra in the principal gradation 
(see \cite{guilherme} for a review). 

Another peculiar feature of integrable models is the existence of B\"acklund Transformations  which relate two 
different  field configurations of certain  non-linear differential equation.  
These B\"acklund transformations, among other applications,  generate an infinite sequence of soliton solutions  
from a non-linear superposition principle (see  \cite{rogers} for review).

 B\"acklund transformations have also been employed to describe  integrable defects \cite{cor-zamb2} in the sense that two solutions of an integrable model may be interpolated  
 by a defect at certain  spatial position.  
 The B\"acklund transformation  connecting the two  field configurations is the key ingredient  to preserve  the integrability of the system.  
 Under such formulation the canonical energy and momentum are no longer conserved  and modifications to ensure its conservation  are required 
 in order to take into account the contribution of the defect \cite{bcz}.
  Well known (relativistic) integrable models as the sine (sinh)-Gordon, Tzitzeica \cite{cor-zamb2}, Lund-Regge \cite{bow} 
and other  (non
relativistic)  models  as Non-Linear Schroedinger (NLS), mKdV, etc  have been studied
within such context \cite{cor-zamb1} .

In ref. \cite{ana-1} we  have constructed  B\"acklund  transformation to the mKdV hierarchy  by  assuming that two field configurations of
the same equation of motion  were related by a gauge transformation and dubbed from now on  B\"acklund-gauge transformation.  As a result  it was subsequently  shown that such 
B\"acklund-gauge transformation acted in a universal manner in all members of the hierarchy, i.e.,  the same   for all models within the mKdV hierarchy.

In this paper we  extend  the results of  \cite{ana-1}  to the  KdV hierarchy. We first  consider  the transformation  from the principal to the 
homogeneous gradation by a global  gauge transformation $g_1$.  Next, by a second, local  gauge transformation $g_2(v, \eps) $, we  realize the 
Miura transformation relating the fields $v$ of the mKdV to those $J_{\eps} =\eps \pa_x v - v^2, \eps = \pm 1$ of the KdV hierarchies. 
In fact, the Miura  transformation displays a sign  ambiguity  represented  by the $\eps$-factor, i.e.,  solutions of the mKdV hierarchy  generate  
two  towers of solutions of the KdV hierarchy labeled by the $\eps$-sign.   
%One of the main results of this paper is to  show that solutions for 
%Backlund equations of the KdV hierarchy are  solved  by mixed Miura mapping  the mKdV  Backlund solutions of opposite $\eps$-signs.

In Section 2 we review the algebraic construction of mKdV hierarchy  and  construct explicitly  
the Miura-gauge transformation  that  maps the equations of the mKdV into   equations of the KdV hierarchy.
We   display the first few positive grade time evolution equations for both hierarchies. 
 In Section 3, from the B\"acklund-gauge transformation constructed in ref \cite{ana-1} we  discuss  
 its Miura extension  and propose B\"acklund-gauge transformation for  the KdV hierarchy.  
 This, however  displays  a sign ambiguity  inherited  from the Miura transformation.  
 Each solution, $v(x, t_N)$ of  the mKdV  hierarchy is mapped into two solutions, $J_{\eps} (x, t_N)= \eps \pa_x v - v^2, \quad \eps=\pm 1 $  of the KdV hierarchy. 
 One of the main    results  of our construction   is to 
  show that a pair of B\"acklund solutions of the mKdV hierarchy, $v_1, v_2$  have to be  mapped 
  into a pair of KdV B\"acklund solutions  of opposite $ \eps$- signs, $J_{1+}=\pa_x v_1 -v_1^2, \quad J_{2-}=-\pa_x v_2 - v_2^2$ in order to satisfy the KdV B\"acklund equations.

In section 4 we discuss the composition  of gauge-B\"acklund transformations  for both hierarchies.  
In fact we start by considering two consecutive gauge-B\"acklund transformations  for the mKdV hierarchy, $K_1$ and $K_2$ and we show that the product,  
$K^{II} = K_2 K_1$  generate the  so called Type-II  B\"acklund transformation proposed in \cite{cor-zamb2}.  We follow the same philosophy  to construct  the Type-II B\"acklund transformation for the KdV hierarchy. 
First by direct  product composition  of two Type-I, $ \tilde K^{II} = \tilde K_2 \tilde K_1$ and  the second by Miura-gauge transforming the Type-II B\"acklund transformation 
of the mKdV hierarchy.  Consistency  is shown that the  two alternative constructions indeed agree.

\section{The Algebraic  Formalism  for KdV and mKdV Hierarchies}

Following the algebraic formalism  described in  ref \cite{ana-1}  we recall that the   
nonline-ar equations of the mKdV hierarchy can be derived from the zero curvature representation,
\br
\lb \pa_x + A_x, \pa_{t_{N}} + A_{t_N} \rb = 0.
\label{zccmkdv}
\er
{ 
	underlined by an affine $\hat {sl}(2)$ 
	centerless Kac-Moody algebra   generated by
	$h^{(m)} = \l^m h,\;\;  E_{\pm \a}^{(m)} = \l^m E_{\pm \a}, \quad \l\in C, \quad m\in Z$ satisfying 
	\br
	[h^{(m)}, E_{\pm \a}^{(n)}]= \pm 2 E_{\pm \a}^{(m+n)}, \quad [E_{\a}^{(m)}, E_{-\a}^{(n)}] = h^{(m+n)}.
	\er
}
Here $ A_x = E^{(1)} + A_0$,  $A_{t_N} =      D^{(N)} + D^{(N-1)} 
+\cdots +D^{(0)}$,  $ E^{(1)}= E_{\a} + \l E_{-\a}$ and  $ A_0 = v(x,t_N) h$\footnote{ $ h=\begin{pmatrix}
	1 & 0\\
	0 & -1
	\end{pmatrix} $, $ E_\a=\begin{pmatrix}
	0 & 1\\
	0 & 0
	\end{pmatrix} $ and $E_{-\a}=\begin{pmatrix}
	0 & 0\\
	1 & 0
	\end{pmatrix}   $}   are constructed according to
the principal gradation, $Q_p= 2 \l {{d \o {d\l}}} + 1/2 h$  which decomposes the affine $\hat {sl}(2) = \sum_{i} \lie_i$  algebra into 
graded subspaces according to powers of the spectral parameter $\l$, ,
\br 
\lb Q_p, \lie_a \rb =a\lie_a, \qquad  \lb \lie_a , \lie_b \rb \in \lie_{a+b}, \qquad  a,b \in Z,
\er
{  where the subspaces $\lie_{2m}$ and $\lie_{2m+1}$ contains the following generators}
\br
\lie_{2m} &=& \{ h^{(m)} = \l^m h\}, \nonu \\
\lie_{2m+1} &=& \{\l^m\(E_{\a} + \l E_{-\a}\), \; \l^m\(E_{\a} - \l E_{-\a}\)\}.
\label{2} 
\er
Here $D^{(i)} \in \lie_i$.  The zero curvature (\ref{zccmkdv}) decomposes according to the graded structure  as
\br
\lb E^{(1)}, D^{(N)}\rb &=0&  \nonu \\
\lb E^{(1)}, D^{(N-1)}\rb + \lb A_0, D^{(N)}\rb +\pa_x D^{(N)} &=0& \nonu \\
\vdots &=& \vdots \nonu \\
\lb A_0, D^{(0)}\rb + \pa_x D^{(0)} - \pa_{t_{N}}A_0 &=&0,
\label{6}
\er
which allows solving  for $D^{(i)}, i=0, \cdots N$ and    the last eqn. in (\ref{6}) yields the time evolution for fields $A_0$. {  We should point out that $D^{(i)}$ are constructed systematically 
	for each value of $N$ and so is $A_{t_N}$.}
The first few explicit  equations are (see for instance \cite{ana-1}, \cite{ana-2}),
\br
4\pa_{t_3} v &-& \pa_x \(\pa_x^2v -2 v^3 \) =0 \qquad  mKdV  \nonu \\
\label{t3} \\
\nonu \\
 16\pa_{t_5}v  &-& \pa_x \(   \pa_x^4v -10v^2(\pa_x^2 v) -10v(\pa_x v)^2+6v^5 \) =0, \nonu \\
\label{t5} \\
\nonu \\
 64\pa_{t_7}v &-& \pa_x \(  \partial_{x}^6v-70 (\partial_{x}v)^{2}(\partial_{x}^2v)-42v(\partial_{x}^2v)^{2}-56 v(\partial_{x}v)(\partial_{x}^3v)\)\nonu \\  
& +&\pa_x \(14 v^{2}\pa_x^4v -140 v^{3}(\partial_{x}v)^{2}-70 v^{4}(\partial_{x}^2v)+20v^{7}  \) =0 \nonu \\ 
\label{t7} \\
  \cdots etc. \nonu 
\er 
Consider now the {\it  global} gauge transformation generated by
\br
g_1 =  \left( \begin{array}{cc}
\zeta & 1  \\
\zeta  & -1  \\
 \end{array} \right) ,   \quad \zeta^2 = \l
 \label{g1}
 \er
 which transforms $ 
A^{princ}_{x,mKdV}  =   E^{(1)} + v(x, t_N) h = \left( \begin{array}{cc}
v & 1  \\
\lambda & -v \\
 \end{array} \right),  
 %\label{aprinc}
$
into
\br 
A^{hom}_{x, mKdV} = g_1 \( A^{princ}_{x, mKdV} \)  g_1^{-1} =  g_1 \( E^{(1)} + v(x, t_N) h \) g_1^{-1} = \left( \begin{array}{cc}
\zeta & v  \\
v & -\zeta \\
 \end{array} \right),  
 \label{ahom}
 \er
 i.e.,  transforms the principal into homogeneous gradation, $Q_{hom} = \zeta {{d \over {d\zeta}}}$ .
 
 A subsequent {\it  local}  Miura-gauge transformation \cite{fukuyama}, \cite{gsf}
 \br
g_2(v, \eps) =  \left( \begin{array}{cc}
1 & \eps  \\
-\eps v  & -v +2 \eps  \zeta  \\
 \end{array} \right) ,  
 \label{g2}
 \er
 transforms $A_{x, mKdV} \rightarrow A_{x, KdV}$. i.e.,
 \br
 A_{x, KdV} = g_2(v, \eps) A^{hom}_{x, mKdV}  g_2^{-1}(v, \eps) -\pa_x g_2(v, \eps) g_2^{-1}(v, \eps) = \left( \begin{array}{cc}
\zeta & -1  \\
J & -\zeta \\
 \end{array} \right) \hspace{0.1cm}\
 \label {g2-m}
 \er
 and realizes the Miura transformation, 
 \br J = \eps \pa_x v -  v^2, \quad \eps^2 =1. \label{miu}
 \er
 We should emphasize  that for each solution $v(x,t_N)$ of the evolution equations for the  mKdV hierarchy,  the Miura transformation (\ref{miu}) 
 generates two towers of solutions, $J_{\eps}(x,t_N), \quad \eps = \pm 1 $,  of the KdV hierarchy \cite{gsf}.
 The zero curvature under the homogeneous gradation \footnote{ Notice that under the  homogeneous 
 gradation the  decomposition of the affine Lie algebra $\hat sl(2) = \sum_a \tilde{\lie}_a, 
 \quad a \in Z, \quad \tilde{\lie}_a = \{ \zeta^a h, \zeta^a E_{\alpha}, \zeta^a E_{-\alpha} \} $}
 \br
\lb \pa_x + A_{x, KdV}, \quad \pa_{t_{N}} + A_{t_N, KdV}  \rb = 0,
\label{zcckdv}
\er
 with $ A_{t_N, KdV} = {\cal {\tilde{D} }} ^{(N)} + {\cal {\tilde{D} }}^{(N-1)} 
+\cdots +{\cal {\tilde{D} }}^{(0)}, \quad {\cal {\tilde{D} }} ^{(j)} \in {\tilde{\lie}} ^{j}$ yields  the KdV hierarchy  equations of motion.  
 For instance
 \br
 A_{t_3, KdV}=\begin{bmatrix}
     \zeta^3+\frac{1}{2}\zeta J+\frac{1}{4}\pa_x J & -\zeta^2-\frac{1}{2}J\\
     \zeta^2J+\frac{1}{2}\zeta\pa_x J+\frac{1}{4}\pa^2_x J+\frac{1}{2}J^2 &-\zeta^3-\frac{1}{2}\zeta J-\frac{1}{4}\pa_x J
 \end{bmatrix}
 \label{a3}
 \er 
 yields the KdV equation
 \br
4\pa_{t_{3}}J -\pa^3_xJ - 6 J\pa_xJ = \(\eps \pa_x -2v\) [4\pa_{t_3}v -\pa_x( \pa^2_x v -2v^3)] = 0,\label{k3} \\ \nonu
\er
Similarly  from  $A_{t_5}$ and  for $A_{t_7}$, given in the appendix,  we find  the Sawada-Kotera equation \cite{sawada-kotera}
\br
& & 16\pa_{t_5}J - \pa^5_xJ-20\pa_xJ \pa^2_xJ -10J \pa^3_xJ -30J^2 \pa_xJ \nonu \\
&=& \(\eps \pa_x -2v\) [16\pa_{t_5}v - \pa_x (\pa^4_{x}v- 10v^2 \pa^2_{x}v -10v (\pa_xv)^2+6v^5    )]= 0, \nonu \\
 \label{k5} 
\er
and   {\footnote{ In general, we find $ KdV (J) = (\eps \pa_x -2v) mKdV (v)$.}}
\br
 & &64\pa_{t_7}J - \pa^7_xJ -70\pa^2_xJ \pa^3_xJ -42 \pa_xJ \pa^4_xJ - 70  (\pa_xJ)^3 
 - 14 J \pa^5_xJ \nonu \\ &-& 280 J \pa_xJ \pa^2_xJ - 70 J^2 \pa^3_xJ - 140J^3 \pa_xJ  \nonu \\
 &=& \( \eps \pa_x -2v\)(64\pa_{t_7}v- \pa_x ( \pa^6_x v - 70 (\pa_xv)^{2}\pa^2_x v - 42v (\pa^2_x v)^{2} - 56 v \pa_xv \pa^3_x v  \nonu \\
 &-&14 v^{2}\pa^4_x v + 140 v^{3}(\pa_xv)^{2}+ 70 v^{4}\pa^2_x v -20v^{7}  ) ) = 0 
\label{k7}
\er
 respectively.  { Eqns.( \ref{k3}-\ref{k7}) are displayed as explicit examples as illustration of the formalism.
 	Higher flows  (time evolutions) can be  systematically constructed for generic $N$ from the same formalism.}

 \section{ B\"acklund Transformation } 
 \subsection{mKdV}
 In this section we start by noticing  that the zero curvature representation (\ref{zccmkdv})  and (\ref{zcckdv})
are invariant under gauge transformations of the type
\br
A_{\mu} (\phi, \pa_x \phi, \cdots ) \rightarrow \tilde {A}_{\mu}= K^{-1} A_{\mu} K + K^{-1} \pa_{\mu}K,
\label{gt}
\er
\noindent
where $ A_\mu $ stands for either $ A_{t_N} $ or $ A_x $.

The key ingredient of this section is  to consider two field configurations $\phi_1$ and $\phi_2$  embedded in $A_{\mu} (\phi_1)$ and $  A_{\mu}(\phi_2) $ satisfying the zero curvature representation
 and  assume that they are related by   a B\"acklund-gauge transformation generated by  $K(\phi_1, \phi_2)$  
preserving the equations of motion 
(e.g,, zero curvature (\ref{zccmkdv})  or (\ref{zcckdv}) ) , i.e., 
\br
K (\phi_1, \phi_2)A_{\mu}(\phi_1)  = A_{\mu}(\phi_2) K(\phi_1, \phi_2) + \pa_{\mu} K(\phi_1, \phi_2).
\label{back}
\er
If we now consider  the Lax operator $L= \pa_x + A_x$ for mKdV case  within the principal gradation,
\br
 {A_x}_{,mKdV} = E^{(1)} + A_0 = \left[ \begin{array}{cc}  \pa_x \phi(x,t_N) & 1 \\ \l & -\pa_x \phi(x, t_N) \end{array} \right]
\er
  is {\it common to all members} of the hierarchy  defined by (\ref{zccmkdv}).
We find that 
\br
K (\phi_1, \phi_2){A_{x}}_{,mKdV}(\phi_1)  ={ A_{x}}_{,mKdV}(\phi_2) K(\phi_1, \phi_2) + \pa_{x} K(\phi_1, \phi_2),
\label{back-mkdv}
\er
where the B\"acklund-gauge generator  $K(\phi_1, \phi_2)$ is given by \cite{ana-1}, \cite{ana-2}
\br
K(\phi_1, \phi_2) = \left[ \begin{array}{cc}  1 & -{{\b}\over {2\l}}e^{-(\phi_1+\phi_2)} \\ -{{\b}\over {2}}e^{(\phi_1+\phi_2)} & 1 \end{array} \right]
\label{gt}
\er
and $\b$ is the B\"acklund parameter.
Eqn.  (\ref{back-mkdv})  is satisfied provided
\br
\pa_x \( \phi_1 - \phi_2\) = -\b \sinh \( \phi_1 + \phi_2 \).
\label{xbl-sg}
\er 
For the sinh-Gordon (s-g) model, the equations of motion $ \pa_{t} \pa_x \phi_a = 2 \sinh 2 \phi_a, \; a=1,2$ 
are satisfied if we further introduce the time component of the B\"acklund transformation,  
\br
\pa_{t} \( \phi_1 + \phi_2\) = {{4}\over {\b}} \sinh \( \phi_2 - \phi_1 \).
\label{tbl-sg}
\er 
Eqn. (\ref{tbl-sg}) is compatible with (\ref{back}) for $A_\mu =A_{t_N}$ with  
\br
 {A_t}_{, s-g} = \left[ \begin{array}{cc}  0 &\l^{-1}  e^{-2\phi} \\  e^{2\phi}  & 0 \end{array} \right].
\er

For  higher graded  time evolutions the time component of the B\"acklund transformation  can be derived  from the appropriated  
time component of the two dimensional gauge potential.  
Several explicit examples within the positive and negative graded  mKdV sub-hierarchies were discussed in \cite{ana-1}. { We now 
give a general argument that the B\"acklund Transformation  derived  from the gauge transformation (\ref{back-mkdv}) 
	for arbitrary $N$ provides equations compatible with the eqn. of motion.
	Consider the zero curvature representation  for certain field configuration, namely $\phi_1$, i.e.,
	\br
	\lb \pa_x + A_x(\phi_1), \pa_{t_{N}} + A_{t_N}(\phi_1) \rb = 0.
	\label{zcc1}
	\er
	Under the gauge transformation,
	\br
	&&K(\phi_1,\phi_2)\lb \pa_x + A_x(\phi_1), \pa_{t_{N}} + A_{t_N}(\phi_1) \rb K(\phi_1,\phi_2)^{-1} \nonu \\
	&&=\lb K( \pa_x + A_x(\phi_1))K^{-1}, K(\pa_{t_{N}} + A_{t_N}(\phi_1))K^{-1} \rb    \nonu \\
	&&=\lb \pa_x + A_x(\phi_2), \pa_{t_{N}} + A_{t_N}(\phi_2) \rb = 0.
	\label{zcc2}
	\er
	where the last equality comes from our assumption  (\ref{back-mkdv}).
	
	The gauge transformation of the  first entry in the zero curvature representation implies the x-component
	of the B\"acklund  transformation (\ref{xbl-sg}).  Since the zero curvature (\ref{zcc1}) 
	and (\ref{zcc2}) implies that both $\phi_1$ and $\phi_2$ satisfy
	the same equation of motion, the gauge transformation (\ref{back-mkdv}) for $A_\mu = A_{t_N}$ of the second entry in (\ref{zcc2}) 
	generates the time component of BT which, by construction has to be consistent with  the equations of motion  with respect to 
	time $t_N$.}

\subsection{KdV}	
	
In order to extend the same  philosophy  to the KdV hierarchy recall the fact   that the two dimensional gauge potential $A_{x,KdV}$   can be 
obtained  by Miura-gauge transformation from the homogeneous mKdV  gauge potentials $ A^{hom}_{mKdV}  $  as in (\ref{g2-m}), i.e.,
\br
 A_{x, KdV} (J)= g_2(v, \eps) g_1 \( A_{x,mKdV}(v) \)  g_1^{-1}  g_2^{-1}(v, \eps) -\pa_x g_2(v, \eps) g_2^{-1}(v, \eps),
\label{xx}
\er
\noindent
  where $v= \pa_x \phi(x,t_N)$
  By assuming   (\ref{back})   for the KdV hierarchy, i.e.,
 \br
\tilde K (J_1, J_2)A_{x,KdV}(J_1)  = A_{x,KdV}(J_2)\tilde  K(J_1, J_2) + \pa_{x} \tilde K(J_1, J_2).
\label{back-tilde}
\er
 the B\"acklund-gauge  transformation  for the KdV hierarchy $ \tilde K(J_1, J_2)$
  cons-tructed in terms of  $K(\phi_1, \phi_2)$   can be written as
\br 
\tilde K = g_2(v_2, \eps_2) \( g_1 K(\phi_1, \phi_2) g_1^{-1}\)  g_2 (v_1, \eps_1)^{-1}.
\label{ktilde}
\er
 At this stage we should recall that for each solution of the mKdV  hierarchy $v$,  the Miura transformation (\ref{miu}) generates two solutions, 
  $J_{\eps_i}= \eps_i \pa_x v_i -  v_i^2, \quad \eps = \pm 1$    satisfying the associated equation of motion of the KdV  hierarchy.  
  This is  precisely  why we assume $\eps_1$ and $\eps_2$ in eqn. (\ref{ktilde}) independent.
 In terms of mKdV variables $v_i = \pa_x \phi_i$, $\tilde K$ 
 %is  given by
 %\br
%\tilde K (\eps_1, \eps_2)  =  \left( \begin{array}{cc}
%X & Y  \\
%Z  & T  \\
 %\end{array} \right) ,  
 %\label{tildeK}
 %\er
 is given for 
 the particular case  where $\eps_1 =-\eps_2\equiv\eps$ and denote $ \tilde K(J_1, J_2) = \tilde K (\eps_1 =-\eps_2\equiv\eps) $ { \bf {\footnote {Notice that $\tilde K$ is given in terms of mKdV variables 
 			$v_1,v_2$ and we need to rewrite it in terms of KdV variables $J_1, J_2$.  
 			This requires  solving Riccati eqn. $v= v(J)$ (\ref{miu}) }}}, 
  \begin{align}
  	& { \tilde {K}(J_1,J_2) }_{11}=1-\frac{v_1\eps }{\zeta}-\frac{\b}{4\zeta}(1-\eps)e^{-p}-\frac{\b}{4\zeta}(1+\eps)e^{p}\nonu\\
  	& { \tilde {K}(J_1,J_2) }_{12}=-\frac{1}{\zeta}\nonu\\
  	& { \tilde {K}(J_1,J_2) }_{21}=\frac{\b}{4\zeta}\left(-v_1(1+\eps)+v_2(1-\eps)\right)e^{-p}\nonu\\
  	& \hspace{2cm}+\frac{\b}{4\zeta}\left(v_1(1-\eps)-v_2(1+\eps)\right)e^{p}-\frac{v_1v_2}{\zeta}\nonu\\
  	& { \tilde {K}(J_1,J_2) }_{22}=-1-\frac{v_2\eps }{\zeta}-\frac{\b}{4\zeta}(1+\eps)e^{-p}-\frac{\b}{4\zeta}(1-\eps)e^{p}\label{kk}
  \end{align}
  where $p= \phi_1+ \phi_2$. 
 Substituting in eqn.  (\ref{back-tilde})  we find the following equations:\
 
 \
 
 \begin{itemize}
 	\item Matrix element 11:
 \end{itemize}
  \br
\zeta^{-1} & :& J_1-\eps \pa_x v_1-\frac{1}{2}\b v_1 (e^p-e^{-p})+v_1v_2=0
 \label{11}
 \er
 \begin{itemize}
 	\item Matrix element 12:
 	\end{itemize}
 \br
 \zeta^{-1} & :&v_1-v_2 + {{\b}\over {2}}(e^p - e^{-p}) =0
 \label{12}
 \er
  \begin{itemize}
  	\item Matrix element 21:
  \end{itemize}
 
 \begin{align}
 &\zeta^0: \hspace{0.3cm} J_1+J_2+\frac{\b v_1}{2}(1+\eps)e^{-p}-\frac{\b v_1}{2}(1-\eps)e^{p}\nonu\\
 &  \hspace{0.5cm} -\frac{\b v_2}{2}(1-\eps)e^{-p}+\frac{\b v_2}{2}(1+\eps)e^{p}+2 v_1 v_2=0\label{jj1} 
 \end{align} 
\begin{align}
 & \zeta^{-1}: \hspace{0.3cm}\eps(J_1v_2-J_2v_1)-v_1\pa_xv_2-v_2\pa_xv1\nonumber\\ & \hspace{1cm}-\frac{\eps\b}{2}v_1v_2(e^p-e^{-p})=0
 \label{21}
\end{align}

 \begin{itemize} 
\item  Matrix element 22:
\end{itemize} 
 
 \br
  \zeta^{-1} \hspace{0.3cm}	-J_2-\eps\pa_xv_2-\frac{\b v_2}{2}(e^p-e^{-p})=0.
 \label{22}
 \er\
 
 \
  Using the mixed Miura transformation, i.e., $\eps_2=-\eps_1\equiv\eps$,   
 \br
J_1 = \eps\pa_x v_1 -v_1^2, \qquad J_2 = -\eps\pa_x v_2 -v_2^2
 \label{mj}
 \er together with the mKdV B\"acklund transformation (\ref{xbl-sg})
  \br
 v_1-v_2 = -{{\b}\over {2}} (e^p - e^{-p}),
 \label{bk}
 \er
 we find that  eqns. (\ref{11}), (\ref{12}), (\ref{21}) and (\ref{22}) are  identically satisfied.  Defining the new variable $Q$ and taking into account 
 the  B\"acklund eqn. (\ref{bk}) we find the following equality
 \br
  \frac{1}{2}Q &=& {\eps v_1}+\frac{\b}{4}(1-\eps)e^{-p}+\frac{\b}{4}(1+\eps)e^{p}\label{halfQ1} \\
  &=& \eps {v_2}+\frac{\b}{4}(1+\eps)e^{-p}+\frac{\b}{4}(1-\eps)e^{p},\label{halfQ2}
 \er
Eliminating $ v_1 $ and $ v_2 $ from eqns \eqref{halfQ1}  and \eqref{halfQ2} we find
 \begin{align}
 & v_1=\frac{\eps}{2}Q-\frac{\b}{4}(\eps-1)e^{-p}-\frac{\b}{4}(\eps+1)e^{p}\label{eqv1}\\
 &  v_2=\frac{\eps}{2}Q-\frac{\b}{4}(\eps+1)e^{-p}-\frac{\b}{4}(\eps-1)e^{p}\label{eqv2}
 \end{align}
 and henceforth
 \begin{align}
 & \frac{\b}{4}\left(-v_1(1+ \eps)+v_2(1-\eps)\right)e^{-p}\nonu\\
 &+ \frac{\b}{4}\left(v_1(1-\eps)-v_2(1+\eps)\right)e^{p}
 -v_1v_2=\frac{\b^2}{4}-\frac{Q^2}{4}.
 \end{align}
 Eqn. (\ref{jj1})  then becomes
 \br
 J_1 + J_2 = {{\b^2 }\over {2}} -  {{Q^2}\over {2}}.
 \er
 From \eqref{halfQ1} and \eqref{halfQ2} we find that 
\br
Q=\eps(v_1+v_2)+\frac{\b}{2}(e^p+e^{-p})
\label{Q}
\er
 Acting with $\pa_x$ in (\ref{Q}) and using (\ref{mj}) and (\ref{bk}),
\begin{align}
\pa_xQ & =\eps\pa_x(v_1+v_2)+\frac{\b}{2}(v_1+v_2)(e^p-e^{-p})\nonu\\
& =\eps\pa_x(v_1+v_2)-(v_1-v_2)(v_1+v_2)\nonu\\
& = J_1-J_2\nonu\\
& = \pa_x \left(\omega_1-\omega_2\right)
\end{align}
where we have used $  J_i \equiv \pa_x w_i, i=1,2 $.  It therefore  follows   that 
 \br
  Q = w_1 - w_2
  \label{ww}
  \er
 and the B\"acklund transformation  for the spatial component of the KdV equation becomes,
 \br
 J_1+J_2  = \pa_x P = {{\b^2 }\over {2}} -  {{(w_1-w_2)^2}\over {2}}, \qquad P = w_1 + w_2.
 \label{3.26}
 \er
 which is in agreement with the B\"acklund transformation  proposed in \cite{eastbrook} and with \cite{villani}.
 
 In the new variable $Q$ defined in \eqref{halfQ1} and \eqref{halfQ2}  we rewrite the gauge-B\"acklund transformation $\tilde K (J_1,J_2)$  in (\ref{kk}) as 
 \br
  \tilde K (J_1,J_2, \b)  = - \frac{1}{\zeta}\left( \begin{array}{cc}
  - \zeta +{{1}\over {2} }Q & 1  \\
{{-\b^2}\over {4}}+{{1}\over {4}}Q^2  & \zeta +{{1}\over {2}}Q \\
 \end{array} \right),                                                   
\label{kkk}
 \er
 Other cases with $\eps_1 = \eps_2 = \pm 1$  lead to trivial B\"acklund transformations in the 
 sense that (\ref{back-tilde}) for $ \tilde K (\pm 1,\pm 1)$  is trivially satisfied   for  mKdV  B\"acklund   and Miura transformations (\ref{xbl-sg}) and (\ref{miu}).  
 There  is no new equation relating the two KdV fields $J_1$ and $J_2$.  From now on   we shall only consider $\tilde K (+1,-1)\equiv \tilde K $ given  in (\ref{kkk}) and
 Miura transformation given by (\ref{mj})

 We now discuss the extension of the B\"acklund transformation to the time component  of the KdV hierarchy.
 Notice that in the zero curvature representation the  spatial component of the two dimensional gauge potential $A_x$ is the same for all flows and therefore 
 universal among the different  evolution equations.  
 They differ  from the time component $A_{t_{N}}$ written  according  to  the  algebraic graded structure and parametrized by the integer $N$.
 \br 
A_{t_N, KdV} = {\cal {\tilde{D} }} ^{(N)} + {\cal {\tilde{D} }}^{(N-1)} 
+\cdots +{\cal {\tilde{D} }}^{(0)}, \quad {\cal {\tilde{D} }} ^{(j)} \in {\tilde{\lie}} ^{j}.
\label{atN}
\er
The  B\"acklund-gauge transformation (\ref{kkk}) acting on the potentials  $A_{t_3, KdV}$, $A_{t_5, KdV}$ and $A_{t_7, KdV}$ given by eqns. (\ref{tt3})- (\ref{tt7}) of the appendix 
leads  to the following  B\"acklund  equations respectively
\br
4\pa_{t_3} P  &=& -Q\pa^2_{x} Q + {{1}\over {2}}\( (\pa_x Q)^2   + 3 (\pa_x P)^2 \) 
\label{t3} \\
\nonu \\
 16\pa_{t_5}P   &=&  -Q \pa^4_{x}Q + \pa_x Q \pa^3_{x}Q + 5 \pa_x P \pa^3_{x}P \nonu \\
 &+&  {{1}\over {2}}\( 5(\pa^2_{x} P)^2   -  (\pa^2_{x} Q)^2 \) + {{5}\over {2}}\pa_x P \( (\pa_x P)^2   + 3 (\pa_x Q)^2 \) 
\label{t5} \\
\nonu \\
 64\pa_{t_7}P &=&    -Q \pa^6_{x}Q + \pa_x Q \pa^5_{x}Q + 7 \pa_x P \pa^5_{x}P  - \pa^2_{x}Q \pa^4_{x}Q + 14 \pa^2_{x}P \pa^4_{x}P\nonu \\
  &+&  35 \pa_xQ \pa^2_{x}Q \pa^2_{x}P + 35 \pa_x P \pa_xQ \pa^3_{x}Q + {{21}\over {2}}(\pa^3_{x}P)^2 + {{1}\over {2}}(\pa^3_{x}Q)^2
 \nonu \\
 &+& {{35}\over {2}}\pa_x P \( (\pa^2_{x}P)^2 + (\pa^2_{x}Q)^2\) +  {{35}\over {2}}\pa^3_{x} P \( (\pa_{x}P)^2 + (\pa_{x}Q)^2\)  
 \nonu \\
 &+&  {{105}\over {4}} (\pa_{x}P)^2  (\pa_{x}Q)^2  + {{35}\over {8}} (\pa_x P)^4 + {{35}\over {8}} (\pa_x Q)^2,
\label{t7} 
  \er
 where $\partial_P=J_1+J_2$.  Equations (\ref{3.26}) and (\ref{t3})   coincide with the B\"acklund  transformation proposed in \cite{eastbrook} for the KdV equation.  
 Equations (\ref{3.26}) and (\ref{t5})   correspond  to those derived  for the Sawada-Kotera equation 
 in \cite{villani} {\footnote{ Notice that there are typos in   eqn. (45.11) of ref. \cite{villani}}.} 
In the appendix we have checked the consistency  between the  spatial,  (\ref{3.26})  and time components (\ref{t3}) -(\ref{t7}) 
of the B\"acklund transformations for $N=3,5$ and $7$.
By direct calculation, using software Mathematica, we indeed recover the evolution equations (\ref{k3})-(\ref{k7}).{ We would 
like to point out that our method is systematic and provides the B\"acklund transformations for
	arbitrary time evolution in 
	terms of its time component 2-d gauge potential $A_{{t_N}, KdV}$ in terms of graded  subspaces $\tilde {D}^{(i)}, i=0, \cdots N$.
	The examples given above for $t_3,t_5$ and $t_7$ just illustrate the potential of the formalism.
}

 \subsection{Examples}

 \begin{itemize}
  \item  Vacuum - One soliton solution
  
  Consider  $\phi_1=0$ and $\phi_2 =  ln ( {{1+\rho }\over {1-\rho}} ), \quad \rho = e^{2k x+2k^N t_N}, N=3,5,7 $   
  two solutions of the mKdV hierarchy.  The mixed Miura transformation  yields 
 \br
 {J^1_+} = \pa_x^2 \phi_1 - (\pa_x \phi_1 )^2  =0, \qquad  {J^2_-} = -\pa_x^2 \phi_2 - (\pa_x \phi_2 )^2 
 \label{jj}
 \er
 Integrating  to obtain $J = \pa_x w$ we find
 \br
 w_1 = 0, \qquad w_2 = -{{4k}\over {1+ \rho}}+2k
 \label{vac-1sol}
 \er
 Type-I B\"acklund transformation $\pa_x (w_1 + w_2) = {{\b^2}\over {2}} -{{1}\over {2}} (w_1 - w_2 )^2$ is satisfied  by (\ref{vac-1sol})  for $\b =\pm 2k$

 \item Scattering of two One-soliton Solutions
 
 Consider the one-soliton of the mKdV hierarchy given by
 \br
 \phi_i = ln \( {{1+R_i\rho }\over {1-R_i\rho}}\),i=1,2 \qquad \rho = e^{2kx +2k^{N}t_N}, \quad N=3,5,7 \cdots
 \label {one-sol}
 \er
 Miura transformation   generates two  one-soliton solutions of the KdV hierarchy, namely
 \br
 J^1_+ &=& \pa_x^2 \phi_1 - (\pa_x \phi_1 )^2;\\
  J^2_- &=& -\pa_x^2 \phi_2 - (\pa_x \phi_2 )^2;  
 \label{two-one}
 \er
 leading to 
 \br
 w_1 = -{{4k}\over {1+ R_1 \rho}} + 2k, \qquad 
  w_2 = -{{4k}\over {1- R_2 \rho}} + 2k
 \label{two-onew}
 \er
 The Type-I  B\"acklund transformation is satisfied for $R_1 = R_2$.  
 Notice that  although $R_1=R_2$, $J_1$ and $J_2$ correspond to different solutions due to opposite $\eps$-sings in the Miura transformation.
 
 \item One-Soliton  into Two-Soliton Solution
 
 Taking $\phi_1 $ given by the one-soliton solution (\ref{one-sol}) and $\phi_2$ by
 \br
 \phi_2 = ln  \({{1+ \d (\rho_1 -\rho_2) - \rho_1 \rho_2}\over {1- \d (\rho_1 -\rho_2) - \rho_1 \rho_2}}\), \quad \d = {{k_1+k_2}\over {k_1-k_2}}
 \er
 leading to 
 \br
 w_2 = -{{2(k_1^2 -k_2^2)(1+\rho_1)(1+\rho_2)}\over {k_1-k_2 - (k_1+k_2)(\rho_1 - \rho_2)  - (k_1-k_2) \rho_1 \rho_2}}
 \er
 where $\rho_i = e^{2k_ix +2k_i^Nt_N}, i=1,2$ 
 satisfy the Type-I B\"acklund transformation for $\b =\pm  2 k_2$.
 \end{itemize}
 All these verifications were made in the software Mathematica.

 \section{ Fusing and Type-II B\"acklund Transformation}
 
 In this section we shall consider  the composition   of two   gauge-B\"acklund transformations  leading to  the Type-II B\"acklund transformation.  
 Let us consider a situation in which we start with a B\"acklund relation
 transforming  solution  $v_1$ into another solution  $v_0$.    A second subsequent B\"acklund relation transforms $v_0$ into $v_2$.  Such  algebraic relation 
 for the mKdV hierarchy   is described by 
 \br
  K^{II}(v_1, v_0, v_2)     =   K(v_2, v_0)  K(v_0, v_1) 
 \label{type2mkdv}
 \er
 where   $K(v_i, v_j)$ is given in (\ref{gt}) with $\b = \b_{ij}$.  
 %Notice that in this setting there are no direct couplings betweem $v_1$ and $v_2$.  
 It leads to 
 \br
  K^{II}(v_1, v_0, v_2)  =  \left[ \begin{array}{cc}  1+ {{\b_{10}\b_{02}}\over {4 \l}}e^q & {{e^{-\phi_0 }}\over {2\l}}(\b_{01}e^{-\phi_1} + \b_{02}e^{-\phi_2} ) \\ 
  -{{1}\over {2}} e^{\phi_0}(\b_{01}e^{\phi_1} + \b_{02}e^{\phi_2} )  & 1+ {{\b_{10}\b_{02}}\over {4 \l}}e^{-q} \end{array} \right]
 \label{type2mkdva1}
 \er
 where $q= \phi_1 - \phi_2$ and $ \s^2 = - {{4}\over {\b_{10} \b_{02}}} $.  Inserting the following identity
 \br
 (\b_{01}e^{\phi_1} + \b_{02}e^{\phi_2} )(\b_{01}e^{-\phi_1} + \b_{02}e^{-\phi_2} )= \b_{01}\b_{02}(\eta  + e^q + e^{-q})
 \er
 where $ \eta =  {{\b_{10}^2 + \b^2_{02}}\over {\b_{10} \b_{02}}}$. Defining $  \L = -\phi_0  -\ln (\b_{02} e^{-\phi_1} + \b_{01}e^{-\phi_2})-\ln {{\s}\over {4}}$, eqn. (\ref{type2mkdva1}) becomes
 \br
  K^{II}(v_1, v_0, v_2)  =  \left[ \begin{array}{cc}  1- {{1}\over {\s^2 \l}}e^q & {{e^{\Lambda -p}}\over {2\l\s}}(e^q + e^{-q} + \eta ) \\ 
  -{{2}\over {\s}} e^{p-\L} & 1- {{1}\over {\l \s^2}}e^{-q} \end{array} \right].
 \label{type2mkdva}
 \er
 Eqn. (\ref{back-mkdv})  with $ K^{II}(v_1, v_0, v_2) $ leads to the following  B\"acklund equations
 \br
 \pa_x q &=& -{{1}\over {2\s}} e^{\L -p}(e^q +e^{-q} + \eta) - {{2}\over {\s}}e^{p-\L} \label{back1}\\
 \pa_x \L  &=& {{1}\over {2\s}} e^{\L -p}(e^q -e^{-q} ). \label{back2}
 \er
 Eqns. (\ref{back1}) and (\ref{back2}) coincide with the $x$-component of the Type-II B\"acklund transformation proposed for the sine-gordon model in \cite{cor-zamb2}.
 Considering now the time component of the 2-D gauge potential  for $t= t_3$ , (i.e., for the mKdV equation), 
 \br
A_{t_3} 
&=& \l E_{\a} + \l^2E_{-\a} + v\l  h + {{1}\over {2}}(\pa_x v - v^2) E_{\a}
- {{1}\over {2}}(\pa_x v + v^2)\l E_{-\a}\nonu \\
&+& {{1}\over {4}}(\pa_x^2v - 2 v^3)h\nonu \\
\er
 we find from eqn. (\ref{back}),
 \br
 16\s^{3}\pa_{t_{3}}q &=& e^{\L-p}\left(e^{q}+e^{-q}+\eta\right)\left[2 \s^{2} (\pa^{2}_{x}p+\pa^{2}_{x}q)+\s^{2} \left(\pa_{x}p+\pa_{x}q\right)^{2}-8e^{q}\right]+\nonu \\
&+& 4 e^{p-\L}\left[-2 \s^{2} (\pa^{2}_{x}p+\pa^{2}_{x}q)+\s^{2} \left(\pa_{x}p+\pa_{x}q\right)^{2}-8e^{-q}\right]+\nonu \\
&+& 16 \s \pa_{x}p\left(e^{q}+e^{-q}+\eta\right)
\label{1.15}
\er
together with
\br
 4\s \pa_{t_3} \L &=& (v_1^2+\pa_x v_1)e^{\L-q-p} - (v_2^2 + \pa_x v_2)e^{\L+q-p}, \label{1.16}
\er
which is  compatible with equations of motion  for the mKdV model.
  These Type-II B\"acklund  equations (\ref{back1})-(\ref{1.16})  coincide with those derived in detail  in ref. \cite{ana-1} 
 where $x \rightarrow x_+, t \rightarrow x_-$ and was extended to all positive higher graded equation within the mKdV hierarchy {\footnote{ Observe that the Type-II B\"acklund transformation  via gauge transformation was constructed in \cite{thiago} where a solution presented 
 there was chosen to reproduce  the B\"acklund transformation  proposed in \cite{cor-zamb2}.  Here we choose a gauge  transformation solution of \cite{thiago} that reproduces  \cite{ana-2}.}}.
 In the case of the KdV  hierarchy 
 \br
 \tilde K^{type II}(J_1, J_0, J_2)     =   \tilde K(J_2, J_0, \b_{02}) \tilde K(J_0, J_1, \b_{01})
 \label{type2}
 \er
 \noindent
 where 
 \begin{equation*}
 \tilde{K}(J_j,J_i,\b_{ij})=-\frac{1}{\zeta}\begin{bmatrix}
 -\zeta+\frac{1}{2}Q_{ij} & 1\\
 -\frac{\b_{ij}^2}{4}+\frac{1}{4}Q_{ij}^2 & \zeta+\frac{1}{2}Q_{ij}
 \end{bmatrix}.
 \end{equation*}
 Such transformation can be interpreted as an extended B\"acklund transformation dubbed Type-II B\"acklund transformation (see \cite{cor-zamb2}).
 Explicitly we find directly from (\ref{type2})
 \br
 {[ \tilde K^{II} (J_1, J_0, J_2) ]}_{11} &  = & 1- {{1}\over {2\zeta}} Q-{{(\b_+ +\b_-)}\over {2\zeta^2}} + {{Q}\over {8\zeta^2}}(Q + P -2\O ) \nonu \\ 
   {[ \tilde{ K}^{II} (J_1, J_0, J_2)]}_{12} &  = & {{1}\over {2\zeta^2}}Q   \nonu \\
     {[ \tilde {K}^{II} (J_1, J_0, J_2)]}_{22} &  = &1+{{1}\over {2\zeta}}Q -{{(\b_+ -\b_-)}\over {2\zeta^2}} + {{Q}\over {8\zeta^2}}(Q-P+2\O) \nonu \\
    {[ \tilde K^{II} (J_1, J_0, J_2)]}_{21} &  = &  -{{\b_+}\over {4\zeta^2}}Q +  {{\b_-}\over {4\zeta^2}}(P-2\O  )  
    +{{Q}\over {8\zeta^2}}(-\O^2 + \O P - {{P^2}\over {4}}  + {{Q^2}\over {4}})\nonu \\
    &-&\frac{\b_-}{\zeta} + {{Q}\over {4\zeta}}(P-2\O)
  \label{type2mkdva}
 \er
 where $Q = Q_{10}+Q_{02} = w_1-w_2, \quad P= Q_{10}-Q_{02}+2\Omega =w_1+w_2,\quad \O = w_0, \quad 4\b_{\pm} = {{\b_{01}^2 \pm \b_{02}^2}}$ 
 and $Q_{ij} = w_i-w_j$.

 Acting with $ \tilde K^{type II}(J_1, J_0, J_2) $  in (\ref{back-tilde}) we find the B\"acklund transformation for the KdV equation, i.e., 
 \br
 \pa_x Q &=& 2\b_- -{{1}\over {2}}PQ + \O Q, \nonu \\
  \pa_x (2\O+ P)  &=&  2\b_+ - {{1}\over {4}}P^2 - {{1}\over {4}}Q^2 - \O^2 + \O P. 
  \label{4.7}
  \er
  Similarly for the time component gauge potential (\ref{tt3}) we find
  \br
  \pa_{t_3}Q &=& {{1}\over {2}}\pa_x P \pa_xQ + {{1}\over {2}} \pa_x \O \pa_x Q + {{1}\over {4}}Q \pa^2_{x}\O + {{1}\over {4}} \O \pa^2_{x}Q - 
  {{P}\over {8}} \pa^2_{x}Q - {{Q}\over {8}}\pa^2_{x}P \nonu \\
  \pa_{t_3}(2\O+P) &=& {{1}\over {4}}(\pa_x P)^2  + {{1}\over {4}}( \pa_x Q)^2 + {{1}\over {2}}\pa_x P \pa_{x}\O 
  +   (\pa_{x}\O)^2 - 
  {{P}\over {8}} \pa^2_{x}P - {{Q}\over {8}}\pa^2_{x}Q \nonu \\
  &+& {{1}\over {4}}P \pa^2_{x}\O +   {{1}\over {4}}\O \pa^2_{x}P - {{1}\over {2}}\O \pa^2_{x}\O.
  \nonu \\
  \label{4.8}
  \er
 Equations (\ref{4.7}) and (\ref{4.8}) are  compatible  and lead to  the eqns. of motion (\ref{k3}). 
  
 Alternatively in terms of the mKdV  B\"acklund transformation (\ref{ktilde}),  eqn. (\ref{type2})  can be obtained by gauge-Miura transformation, i.e.,
 
 \begin{align*}
  \tilde{K}^{TypeII}(J_1,J_0,J_2)=g_2(v_2,\eps_2)g_1 \left(K(\phi_2,\phi_0)\mathbb{I}K(\phi_0,\phi_1)\right)g_1^{-1}g_2(v_1,\eps_1)^{-1}
 \end{align*}
 where we may introduce the identity element, $ \mathbb{I}=g_1^{-1}g_2(v_0,\eps_0)^{-1}g_2(v_0,\eps_0)g_1 $ depending upon an arbitrary $\eps$-sign, say, $\eps_0$.
 As argued when establishing (\ref{ktilde}), we  are considering transitions with opposite $\eps$-signs such that 
  $ \eps_1=-\eps_0= \eps$  and $\eps_0 = -\eps_2 = -\eps$. It therefore follows that
 \begin{align}
&  \tilde{K}^{TypeII}(J_1,J_0,J_2)= g_2(v_2,\eps)g_1\left[K(\phi_2,\phi_0)K(\phi_0,\phi_1)\right]g_1^{-1}g_2(v_1,\eps)^{-1}\nonu\\
&\hspace{3.1cm} = g_2(v_2,\eps)g_1 K^{II}(\phi_2,\phi_1)g_1^{-1}g_2(v_1,\eps)^{-1}\label{type22}
\end{align}
 The equation \eqref{type22} yields
\begin{align}
& \left[\tilde{K}^{TypeII}\left(J_1,J_0,J_2\right)\right]_{11}=1+\frac{(1+\eps)}{4\s \zeta}e^{\L-p}\left(e^q+e^{-q}+\eta\right)-\frac{(1-\eps)}{\s \zeta}e^{p-\L}\nonu\\
& \hspace{4	cm}-\frac{(1-\eps)}{2\s^2\zeta^2}e^{-q}-\frac{(1+\eps)}{2\s^2\zeta^2}e^{q}-\frac{(1-\eps)}{\s \zeta^2}v_1e^{p-\L}\nonu\\
& \hspace{4cm}-\frac{(1+\eps)}{4\s\zeta^2}v_1e^{\L-p}\left(e^q+e^{-q}+\eta\right)\nonu\\
&\left[\tilde{K}^{TypeII}\left(J_1,J_0,J_2\right)\right]_{12}=\frac{(1-\eps)}{\s\zeta^2}e^{p-\L}-\frac{(1+\eps)}{4\s\zeta^2}e^{\L-p}\left(e^q+e^{-q}+\eta\right)\nonu
\end{align} 
\begin{align} 
& \left[\tilde{K}^{TypeII}\left(J_1,J_0,J_2\right)\right]_{21}=-\frac{\eps}{\s^2\zeta}\left(e^q-e^{-q}\right)-\frac{(1-\eps)}{\s\zeta}(v_1+v_2)e^{p-\L}\nonu\\
&\hspace{4cm}-\frac{(1+\eps)}{4\s\zeta}\left(v_1+v_2\right)e^{\L-p}\left(e^q+e^{-q}+\eta\right)\nonu\\
&\hspace{4cm}+\frac{(v_1+v_2)}{2\s^2\zeta^2}\left(e^q-e^{-q}\right)-\frac{\eps(v_1-v_2)}{2\s^2\zeta^2}\left(e^q+e^{-q}\right)\nonu\\
&\hspace{4cm}+\frac{(1+\eps)}{4\s\zeta^2}v_1v_2e^{\L-p}\left(e^q+e^{-q}+\eta\right)-\frac{(1-\eps)}{\s\zeta^2}v_1v_2e^{p-\L}\nonu\\
& \left[\tilde{K}^{TypeII}\left(J_1,J_0,J_2\right)\right]_{22}=1-\frac{(1+\eps)}{4\s \zeta}e^{\L-p}\left(e^q+e^{-q}+\eta\right)+\frac{(1-\eps)}{\s \zeta}e^{p-\L}\nonu\\
& \hspace{4	cm}-\frac{(1+\eps)}{2\s^2\zeta^2}e^{-q}-\frac{(1-\eps)}{2\s^2\zeta^2}e^{q}+\frac{(1-\eps)}{\s \zeta^2}v_2e^{p-\L}\nonu\\
& \hspace{4cm}+\frac{(1+\eps)}{4\s\zeta^2}v_2e^{\L-p}\left(e^q+e^{-q}+\eta\right)  \label{type2mkdvb}
\end{align}

 Comparing  the matrix elements of  (\ref{type2}) with (\ref{type22}) we find the following relations between the mKdV and KdV variables:\
 
 \
 
 \begin{itemize}
 	\item matrix element 11
 \end{itemize}
 \begin{align}
 & \zeta^{-1}:\hspace{0.6cm}-\frac{1}{2}Q=\frac{(1+\eps)}{4\s }e^{\L-p}\left(e^q+e^{-q}+\eta\right)-\frac{(1-\eps)}{\s}e^{p-\L}\label{4.16}\\
 &\zeta^{-2}:\hspace{0.6cm}-{{(\b_+ +\b_-)}\over {2}} + {{Q}\over {8}}(Q + P -2\O )=-\frac{(1-\eps)}{2\s^2}e^{-q}-\frac{(1+\eps)}{2\s^2\zeta^2}e^{q}\nonu\\
 & \hspace{2.2cm}-\frac{(1-\eps)}{\s\zeta^2}v_1e^{p-\L}-\frac{(1+\eps)}{4\s\zeta^2}v_1e^{\L-p}\left(e^q+e^{-q}+\eta\right)\label{4.17}
 \end{align}
\begin{itemize}
	\item matrix element 21:
\end{itemize}
 \begin{align}
 & \zeta^{-1}:\hspace{0.6cm} -\b_- + {{Q}\over {4}}(P-2\O)=-\frac{\eps}{\s^2}\left(e^q-e^{-q}\right)-\frac{(1-\eps)}{\s}(v_1+v_2)e^{p-\L}\nonu\\
 &\hspace{4cm}-\frac{(1+\eps)}{4\s\zeta}\left(v_1+v_2\right)e^{\L-p}\left(e^q+e^{-q}+\eta\right)\label{4.18}\\
 &\zeta^{-2}:\hspace{0.6cm} -{{\b_+}\over {4}}Q +  {{\b_-}\over {4}}(P-2\O  )  
  +{{Q}\over {8}}(-\O^2 + \O P - {{P^2}\over {4}}  + {{Q^2}\over {4}})=\nonu\\
  & \hspace{2.2cm}  +\frac{(v_1+v_2)}{2\s^2}\left(e^q-e^{-q}\right)-\frac{\eps(v_1-v_2)}{2\s^2}\left(e^q+e^{-q}\right)\nonu\\
  &\hspace{2.2cm}+\frac{(1+\eps)}{4\s}v_1v_2e^{\L-p}\left(e^q+e^{-q}+\eta\right)-\frac{(1-\eps)}{\s}v_1v_2e^{p-\L}\label{4.19}
  \end{align}
  
  \begin{itemize}
  	\item matrix element 22:
  \end{itemize}
  
  \begin{align}
  & \zeta^{-2}:\hspace{0.6cm} -{{(\b_+ -\b_-)}\over {2}} + {{Q}\over {8}}(Q-P+2\O)=-\frac{(1+\eps)}{2\s^2}e^{-q}-\frac{(1-\eps)}{2\s^2}e^{q}\nonu\\
  & \hspace{2.2cm} +\frac{(1-\eps)}{\s }v_2e^{p-\L}+\frac{(1+\eps)}{4\s}v_2e^{\L-p}\left(e^q+e^{-q}+\eta\right). \label{2.20}
  \end{align}
  \noindent
  The element 12 and the element 22 with $ \zeta^{-1} $ gives us the same result of \eqref{4.16}.
 Eliminating the mKdV variables $p,q$ and $\L$  we recover the Type-II B\"acklund transformation for the KdV hierarchy (\ref{4.7}) as shown in the  appendix.
 
 \subsection{Examples  and Solutions}
 \begin{itemize}
  \item Vacuum - 1-soliton - Vacuum
  
The first example is to consider vacuum to 1-soliton  and back to  vacuum  again given by the following configuration,
 \br
 w_1 = 0, \qquad w_0 = \O = -{{4k}\over {1+\rho (x, t_N)}}+2k, \quad w_2 =0
 \label{vac-1-vac}
 \er
 \noindent
 with $ \rho(x,t_N)=e^{2kx+2k^Nt_N} $.
 It is straightforward to check that eqns. (\ref{4.7}) and (\ref{4.8}) are  satisfied for $\b_- = 0$ and $ \b_+=2k^2 $.
 
\item 1-soliton - 2soliton - 1-soliton 

 Consider now  a configuration  of 1-soliton transforming into a 2-solitons solution and back to 1-soliton.  It is described by
 \br
w_i = -{{4k_i}\over {1+\rho_i (x, t_N)}}+ 2k_i, \quad i=1,2 \quad\quad  \rho_i = e^{2k_i x + 2 k_i^N t_N}
\label{1soli}\er
\br
\O = w_0 = -{{2(k_1^2 -k_2^2)(1+\rho_1)(1+\rho_2)}\over {k_1-k_2 - (k_1+k_2)(\rho_1 - \rho_2)  - (k_1-k_2) \rho_1 \rho_2}}
 \label{121}
 \er
 Eqns. (\ref{4.7}) and (\ref{4.8}) are  satisfied for $\b_- = k_2^2 -k_1^2 $  and $\b_+ = k_1^2 +k_2^2 $.
 
 \item Vacuum - 1-soliton - 2-soliton 
 
 Consider  the solution of eqn. (\ref{4.7}) and (\ref{4.8}) 
 \br
 w_1=0, w_0=\O =-{{4k_1}\over {1+\rho_1 (x, t_N)}}+ 2k_1
 \er
 and
 \br
 w_2 = -{{2(k_1^2 -k_2^2)(1+\rho_1)(1+\rho_2)}\over {k_1-k_2 - (k_1+k_2)(\rho_1 - \rho_2)  - (k_1-k_2) \rho_1 \rho_2}},
 \label{012}
 \er
 \noindent where $ \rho_i = e^{2k_i x + 2 k_i^N t_N} $
  Eqns. (\ref{4.7}) and (\ref{4.8}) are  satisfied for $\b_- = k_1^2 -k_2^2 $  and $\b_+ = k_1^2 +k_2^2 $.
 \end{itemize}
 
 \section{Conclusions}

 This paper  follows the  line of reasoning of ref. \cite{ana-1} where we have constructed  B\"acklund transformation for 
 the entire  mKdV hierarchy from an universal  gauge transformation.  Such B\"acklund-gauge transformation  relates two  different field 
 configurations,  preserves  the zero curvature and  henceforth its corresponding  evolution equations.
 
 The main result of this paper is the   extension of  such construction to the KdV hierarchy by proposing a Miura-gauge transformation 
 denoted by the product  $g_2g_1$  given  in (\ref{g1}) and (\ref{g2})
 mapping the mKdV into the KdV hierarchy 
 (see (\ref{g2-m})).
 A subtle point is that such Miura mapping  allows a sign ambiguity such that each solution of the mKdV  hierarchy  defines two solutions for its  KdV counterpart.
 The B\"acklund-gauge transformation for the KdV hierarchy is constructed by Miura-gauge transforming the  B\"acklund transformation  of the mKdV system 
 as shown  in (\ref{ktilde}). 
 An interesting  fact is that the B\"acklund transformation  for the KdV hierarchy is solved by mixed Miura  solutions generated by  the mKdV B\"acklund solutions.
A few simple explicit examples illustrate our conjecture. A more general  evidence of the mixed Miura solutions  is shown to  agree with the B\"acklund 
transformation proposed in \cite{eastbrook} and \cite{villani} for the first two KdV flows.
 
 The composition law of two subsequent   B\"acklund-gauge  transformations  leading  to Type-II  B\"acklund  transformation 
 (see (\ref{type2mkdv})) 
 introduced 
 in \cite{cor-zamb2} in the context of sine-Gordon and Tzitzeica models was  extended to the KdV  hierarchy. 
 Within our construction,  we have employed  the direct fusion of two KdV 
  B\"acklund-gauge  transformations 
  in (\ref{type2}) 
  and alternatively,  the Miura  transformation of mKdV  Type-II  B\"acklund transformation 
  as  shown in (\ref{type22}).  
  These  two approaches  generate  relations
  between the mKdV and KdV variables  which were shown
  in the Appendix 6.2,  to be consistent.

Finally we should mention that the idea of an  universality of the B\"acklund-gauge  transformation is   most probably  
valid for other hierarchies  such as the AKNS  and higher rank Toda theories. It would be interesting to see how  such examples can be worked out technically.

It should be interesting to develop the concept of integrable hierarchies for discrete cases and investigate whether  the arguments involving B\"acklund-gauge transformation 
employed  in this paper  
can be extended.    The  relation between the integrable discrete mKdV  \cite{suris}  and   its Miura transformation to discrete  KdV 
equations 
should be understood under the algebraic formalism.

 \section{Appendix}
 
 \subsection{ Zero Curvature for KdV hierarchy}
 Here we  write down the time component of the  two dimensional gauge potential generating the first three flows for the KdV hierarchy.
 Let 
 \br A_{t_N, KdV} = {\cal {D^{\prime} }} ^{(N)} + {\cal {D^{\prime} }}^{(N-1)} 
+\cdots +{\cal {D^{\prime} }}^{(0)}, \quad {\cal {D^{\prime} }} ^{(j)} \in {\lie^{\prime}} ^{j}
\er
where we find by solving the zero curvature representation (\ref{zcckdv}) in the homogeneous gradation. For $N=3$ we have
\br
& &{\cal {D^{\prime}}}^{(3)}=\zeta^3 h,\nonumber \\
& &{\cal {D^{\prime}}}^{(2)}=-\zeta^2 E_{\a}+J\zeta^2E_{-\a},\nonumber \\
& &{\cal {D^{\prime}}}^{(1)}=\frac{1}{2}\pa_{x}J\zeta E_{-\a}+\frac{1}{2}J\zeta h,\nonumber \\
& &{\cal {D^{\prime}}}^{(0)}=-\frac{1}{2}JE_{\a}+\left(\frac{1}{4}\pa^2_{x}J+\frac{1}{2}J^{2}\right)E_{-\a}+\frac{1}{4}\pa_{x}J h,\nonumber \\
\label{tt3}
\er
and $N=5$ 
\br
& &{\cal {D^{\prime}}}^{(5)}=\zeta^5h,\nonumber\\
& &{\cal {D^{\prime}}}^{(4)}=-\zeta^4 E_{\a}+J\zeta^4 E_{-\a},\nonumber\\
& &{\cal {D^{\prime}}}^{(3)}=\frac{1}{2}\pa_{x}J\zeta^3 E_{-\a}+\frac{1}{2}J\zeta^3 h,\nonumber\\
& &{\cal {D^{\prime}}}^{(2)}=-\frac{1}{2}J\zeta^2E_{\a}+\left(\frac{1}{4}\pa^2_{x}J+\frac{1}{2}J^{2}\right)\zeta^2E_{-\a}\zeta^2+\frac{1}{4}\pa_{x}J \zeta^2 h, \nonumber\\
& &{\cal {D^{\prime}}}^{(1)}=\left(\frac{1}{8}\pa^3_{x}J+\frac{3}{4}J\pa_{x}J\right)\zeta E_{-\a} +\left(\frac{1}{8}\pa^2_{x}J+\frac{3}{8}J^{2}\right)\zeta h \nonumber\\
& &{\cal {D^{\prime}}}^{(0)}=\left(-\frac{1}{8}\pa^2_{x}J-\frac{3}{8}J^{2}\right)E_{\a}+\left(\frac{1}{16}\pa^3_{x}J+\frac{3}{8}J\pa_{x}J\right)h \nonu \\
&+&\left(\frac{1}{16}\pa^4_{x}J 
+\frac{3}{8}(\pa_{x}J)^{2}+\frac{1}{2}J\pa^2_{x}J+\frac{3}{8}J^{3}\right)E_{-\a}   \nonu \\
\label{tt5}
\er
and for $N=7$
\br
& &{\cal{D^{\prime}}}^{(7)}=\zeta^7h,\nonumber\\
& &{\cal{D^{\prime}}}^{(6)}=-\zeta^6E_{\a}+J\zeta^6E_{-\a}, \nonumber\\
& &{\cal{D^{\prime}}}^{(5)}=\frac{1}{2}\pa_{x}J\zeta^5E_{-\a}+\frac{1}{2}J\zeta^5h,\nonumber\\
& &{\cal{ D^{\prime}}}^{(4)}=-\frac{1}{2}J\zeta^4E_{\a}+\left(\frac{1}{4}\pa^2_{x}J+\frac{1}{2}J^{2}\right)\zeta^4E_{-\a}+\frac{1}{4}\pa_{x}J\zeta^4h,\nonumber\\
& &{\cal{ D^{\prime}}}^{(3)}=\left(\frac{1}{8}\pa^3_{x}J+\frac{3}{4}J\pa_{x}J\right)\zeta^3E_{-\a}+\left(\frac{1}{8}\pa^2_{x}J+\frac{3}{8}J^{2}\right)\zeta^3h\nonumber\\
& &{\cal{D^{\prime}}}^{(2)}=\left(-\frac{1}{8}\pa^2_{x}J-\frac{3}{8}J^{2}\right)\zeta^2E_{\a} +\left(\frac{1}{16}\pa^3_{x}J+\frac{3}{8}J\pa_{x}J\right)\zeta^2h \nonu \\
&+&\left(\frac{1}{16}\pa^4_{x}J +\frac{3}{8}(\pa_{x}J)^{2}+\frac{1}{2}J\pa^2_{x}J+\frac{3}{8}J^{3}\right)\zeta^2E_{-\a} \nonumber \\ 
& &{\cal{D^{\prime}}}^{(1)}=\left(\frac{1}{32}\pa^5_{x}J+\frac{5}{8}(\pa_{x}J)(\pa^2_{x}J)+\frac{5}{16}J(\pa^3_{x}J)+\frac{15}{16}J^{2}(\pa_{x}J)\right)\zeta E_{-\a} \nonumber \\
    & +&\left(\frac{1}{32}(\pa^4_{x}J)+\frac{5}{32}(\pa_{x}J)^{2}+\frac{5}{16}J(\pa^2_{x}J)+\frac{5}{16}J^{3}\right)\zeta h\nonumber \\
    &+ & {\cal{D^{\prime}}}^{(0)}=\left(-\frac{1}{32}(\pa^4_{x}J)-\frac{5}{32}(\pa_{x}J)^{2}-\frac{5}{16}J(\pa^2_{x}J)-\frac{5}{16}J^{3}\right)E_{\a}\nonumber \\
    &+&\left(\frac{1}{64}(\pa^6_{x}J)+\frac{5}{16}(\pa^2_{x}J)^{2}+\frac{15}{32}(\pa_{x}J)(\pa^3_{x}J)\right)E_{-\a}+\nonumber \\
    &+&\left(\frac{3}{16}J(\pa^4_{x}J)+\frac{35}{32}J(\pa_{x}J)^{2}+\frac{25}{32}J^{2}(\pa^2_{x}J)+\frac{5}{16}J^{4}\right)E_{-\a}+\nonumber \\
    &+&\left(\frac{1}{64}(\pa^5_{x}J)+\frac{5}{16}(\pa_{x}J)(\pa^2_{x}J)+\frac{5}{32}J(\pa^3_{x}J)+\frac{15}{32}J^{2}(\pa_{x}J)\right)h
    \label{tt7}
\er

\subsection{Equivalence between mKdV and KdV variables }

We now  verify  the equivalence between mKdV and KdV variables
From (\ref{4.16}) we find

\begin{align}
& \pa_xQ=-\frac{(1+\eps)}{2\s}\pa_x\left(\L-p\right)e^{\L-p}(e^q+e^{-q}+\eta)\nonu\\
& \hspace{1.2cm}-\frac{(1+\eps)}{2\s}\pa_xqe^{\L-p}(e^q-e^{-q})+\frac{2(1-\eps)}{\s}\pa_x(p-\L)e^{p-\L}
\label{77} 
\end{align}
and using \eqref{4.18}, \eqref{back1} and\eqref{back2} we obtain \

\br 
\pa_xQ=2\b_--\frac{Q P}{2}+Q\O
\label{6.6}
\er

Consider now $P = J_1+J_2 = \pa_x (w_1+w_2)$.  In terms of Miura transformation
\br
\pa_x P &=& \eps\pa_x (v_1+v_2) - (v_1^2 + v_2^2)
 =\eps\pa_x^2 p-\frac{1}{2}(\pa_x p)^2-\frac{1}{2}(\pa_x q)^2. \label{6.7}) 
\er
Acting with $\pa_x$ in (\ref{4.18}) and using \eqref{6.7} and \eqref{4.16} we find 
\begin{align}
& \frac{Q}{2}\left(\pa_x P+2\pa_x\O\right)=\frac{\pa_xQ}{2}(P-2\O)+\frac{1}{2}Q\left[(\pa_xp)^2-(\pa_xq)^2\right]\nonu\\
& -Q \pa_x p\pa_x \L+\frac{(1+\eps)}{2\s}\pa_x p\pa_xqe^{\L-p}(e^q-e^{-q})+\frac{2\eps}{\s^2}\pa_xq(e^q+e^{-q})
\label{6.8}
\end{align}
where $p=v_1+v_2, \quad q = v_1-v_2$.  Substituting the equation \eqref{4.16} in the equation \eqref{4.19} we find
\begin{align}
& \frac{Q}{2}\left[(\pa_xp)^2-(\pa_xq)^2\right]=\frac{2\pa_x p}{\s^2}(e^q-e^{-q})-\frac{2\eps}{\s^2}\pa_xq(e^q+e^{-q})\nonu\\
& -\frac{Q}{2}\left(-\O^2+\O P-\frac{P^2}{4}+\frac{Q^2}{4}\right)+\b_+Q-\b_-(P-2\O)
\end{align}

Substituting this result in \eqref{6.8} and eliminating the mKdV variables using \eqref{back1} and \eqref{back2}   we obtain
\br
\pa_xP + 2 \pa_x \O = -{{1}\over {4}}(P^2+ Q^2) + \O P - \O^2 + 2 \b_+
\label{6.10}
\er
Eqns. (\ref{6.6}) and (\ref{6.10}) correspond  precisely to the Type-II B\"acklund for the KdV hierarchy.

%%%%%%%%%%%%%%%%%%%%%%%%%%%%%%%%%%%%%%%%%%%%%%%%%%%

\subsection{  Consistency with  Equations of Motion}

In this appendix we verify that the compatibility of B\"acklund transformations lead us to the equation of motion.

We start with the spatial part which is common to all  $ N $. For the KdV equation it is given by

\br
\pa_x P=\frac{\b^2}{2}-\frac{1}{2}Q^2.
\er
\noindent
In what follows it  will be useful calculate its spatial derivatives:
\begin{align}
& \pa^2_x P=-Q\pa_xQ \label{d2};\\
& \pa^3_x P=-(\pa_x Q)^2-Q (\pa^2_x Q)\label{d3};\\
& \pa^4_x P=-3(\pa_x Q)(\pa^2_x Q)-Q(\pa^3_x Q)\label{d4};\\
& \pa^5_x P=-3(\pa^2_x)^2-4(\pa_x Q)(\pa^3_x Q)-Q(\pa^4_x Q)\label{d5};\\
& \pa^6_x P=-10(\pa^2_x Q)(\pa^3_x Q)-5(\pa_x Q)(\pa^4_x Q)-Q(\pa^5_x Q)\label{d6};\\
& \pa^7_x P=-10(\pa^3_x Q)^2-15(\pa^2_x Q)(\pa^4_x Q)-6(\pa_x Q)(\pa^5_x Q)-Q\pa^6_xQ\label{d7};\\
& \pa^8_x P=-35(\pa^3_x Q)(\pa^4_x Q)-21(\pa^2_x Q)(\pa^5_x Q)-7(\pa_x Q)(\pa^6_x Q)-Q(\pa^7_x Q)\label{d8}.
\end{align}

\subsubsection{N=3 (KdV)}

The temporal part of the KdV BT is given by

\br
4 \pa_{t_3}P=-Q(\pa^2_x Q)+\frac{1}{2}\left[(\pa_x Q)^2+(\pa_x P)^2\right].
\label{KdVequation}
\er

In order to verify the consistency of this transformation we act with  the spatial derivative to obtain

\br
4 \pa_x \pa_{t_3}P=-Q\pa^3_x Q+3(\pa_x P)(\pa^2_x P),
\er

\noindent
eliminating  the term $ -Q\pa^3_xQ$   from equation \eqref{d4}  we find

\br
4 \pa_x \pa_{t_3}P=\pa^4_x P+3 (\pa_x P)(\pa^2_x P)+3(\pa_x Q)(\pa^2_x Q).
\label{9.11}
\er
Substituting 
\br \pa_xP= J_1+ J_2, \qquad \pa_xQ = J_1-J_2 \label{pq}
\er
(\ref{9.11}) becomes precisely the sum of two KdV equations.

\subsubsection{N=5 }

The temporal part of the BT for N=5 equation is given by
\begin{align} 
16\pa_{t_5}P=& -Q(\pa^4_x Q)+(\pa_x Q)(\pa^3_x Q)+5(\pa_x P)(\pa^3_x P)+\frac{5}{2}(\pa^2_x P)^2-\frac{1}{2}(\pa^2_x Q)^2\nonu\\
 &+\frac{5}{2}(\pa_x P)\left[(\pa_x P)^2+3(\pa_x Q)^2\right]
 \label{BTN=5}
\end{align}
Acting $ \pa_x $ in the equation \eqref{BTN=5} we obtain
\begin{align}
16\pa_x\pa_{t_5}P=&-Q(\pa^5_x Q)+10(\pa^2_x P)(\pa^3_x P)+5(\pa_x P)(\pa^4_x P)+\frac{15}{2}(\pa_x P)^2(\pa^2_x P)\nonu\\&+\frac{15}{2}(\pa^2_x P)(\pa_x Q)^2+15(\pa_x P)(\pa_x Q)(\pa^2_x Q).
\end{align}
Then we isolate the term $ -Q(\pa^5_x Q) $ from  equation \eqref{d6} to find
\begin{align}
16\pa_x\pa_{t_5}P=& \pa^6_x P+10(\pa^2_x Q)(\pa^3_x Q)+5(\pa_x Q)(\pa^4_x Q)+10(\pa^2_x P)(\pa^3_x P)+5(\pa_x P)(\pa^4_x P)\nonu\\
& +\frac{15}{2}(\pa^2_x P)\left[(\pa_x P)^2+(\pa_x P)^2\right]+15(\pa_x P)(\pa_x Q)(\pa^2_x Q),
\end{align}
Substituting (\ref{pq}) we obtain  the sum of two  equations   for $ N=5 $, i.e., eqn. (\ref{k5}).

\subsubsection{N=7}

The temporal BT for the $ N=7 $ equation is
\begin{align}
64\pa_{t_7}P=&-Q(\pa^6_xQ)+(\pa_xQ)(\pa^5_xQ)+7(\pa_x P)(\pa^5_x P)-(\pa^2_xQ)(\pa^4_xQ)+14(\pa^2_xP)(\pa^4_xP)\nonu \\
& +\frac{1}{2}(\pa^3_xQ)^2+\frac{21}{2}(\pa^3_x P)^2+\frac{35}{2}(\pa^3_x P)(\pa_x P)^2+\frac{35}{2}(\pa^3_x P)(\pa_x Q)^2\nonu\\
& +35(\pa^3_x Q)(\pa_x P)(\pa_x Q)+\frac{35}{2}(\pa_x P)\left[(\pa^2_x P)^2+(\pa^2_x Q)^2\right]+35(\pa^2_x P)(\pa^2_x Q)^2(\pa_x Q)\nonu\\
& +\frac{35}{8}\left[(\pa_x P)^2+(\pa_x Q)^2\right]+\frac{105}{4}(\pa_x Q)^2(\pa_x P)^2.
\end{align}
Likewise we did for other values of $ N $, acting  $ \pa_x $ in the above equation. Then we isolate $ -Q(\pa^7_x Q) $ from  equation \eqref{d8} 
to find
\begin{align}
64\pa_x\pa_{t_7}P=& \pa^8_x P+(\pa_x P)(\pa^6_x P)+7(\pa_x Q)(\pa^6_x Q)+21(\pa^2_x P)(\pa^5_x P)+21(\pa^2_x Q)(\pa^5_x Q)\nonu\\
& +35(\pa^3_x P)(\pa^4_x P)+35(\pa^3_x Q)(\pa^4_x Q)+\frac{35}{2}(\pa^4_x P)\left[(\pa_x P)^2+(\pa_x Q)^2\right]\nonu\\
& +35(\pa_x P)(\pa_x Q)(\pa^4_x Q)+70(\pa_x P)\left[(\pa^2_x P)(\pa^3_x P)+(\pa^2_x Q)(\pa^3_x Q)\right]\nonu\\
& +70(\pa_x Q)\left[(\pa^2_x Q)(\pa^3_x P)+(\pa^2_x P)(\pa^3_x Q)\right]+\frac{35}{2}(\pa^2_xP)^3\nonu\\
& +\frac{105}{2}(\pa^2_x P)(\pa^2_x Q)^2+\frac{35}{2}(\pa^2_x P)(\pa_x P)^3+\frac{35}{2}(\pa^2_x Q)(\pa_x Q)^3\nonu\\
&+\frac{105}{2}(\pa^2_x Q)(\pa_x Q)(\pa_x P)^2+\frac{105}{2}(\pa^2_x P)(\pa_x P)(\pa_x Q)^2
\end{align}
Substituting (\ref{pq}) we obtain 
 the equation of motion for $ N=7 $ (\ref{k7}).

 \vskip 0.5cm
{\bf  Acknowledgments}\\
 AHZ and JFG were partially supported by
CNPq and  Fapesp. ALR was supported by Fapesp under Proc. No. 2015/00025-9. {  We  would like to 
	thank Prof. Wen-li Yang for making  a few pages of the book refered below available to us.}

{  Note Added:   After this paper was finished we were informed about the existence of the book ``Introduction to Soliton Theory'' by Dendyuan Chen, Science Press, 
	Beijing, (in chineese) that  may contain some overlaping results. }

\end{document}